\documentclass[twocolumn, amsmath,superscriptaddress, amsfonts,prb]{revtex4}
\usepackage{graphicx}
\usepackage{url}
\begin{document}

\title{Hot spots spontaneously emerging in thin film photovoltaics}

\author{A. Vasko}\affiliation{Department of Physics and Astronomy, University of Toledo, Toledo, OH 43606, USA}
\author{A. Vijh}\affiliation{Xunlight Corporation, 3145 Nebraska Avenue, Toledo, OH 43607, USA}
\author{V. G. Karpov}\email{victor.karpov@utoledo.edu}\affiliation{Department of Physics and Astronomy, University of Toledo, Toledo, OH 43606, USA}
\begin{abstract}

We present data exhibiting hot spots spontaneously emerging in forward biased thin film photovoltaics based on a-Si:H technology. These spots evolve over time shrinking in their diameter and increasing temperature  up to approximately 300 $^o$C above that of the surrounding area. Our numerical approach explores a system of many identical diodes in parallel connected through the resistive electrode and through thermal connectors,  a model which couples electric and thermal processes. The modeling results show that hot spots emerge collapsing from a rather large area of nonuniform temperature, then collapse to local entities. Finally, we present a simplified analytical treatment establishing relations between the hot spot parameters.

\end{abstract}

\date{\today}

\maketitle
\section{Introduction}\label{sec:intro}

It is generally known to almost everyone in PV community that large area photovoltaic (PV) modules are often laterally nonuniform. The lateral nonuniformity can be especially strong in thin film PV, revealing itself in variations between local PV parameters in different areas of a module or in variations between the parameters of nominally identical solar cells cut from the same module. The nonuniformity causes problems with PV reliability; it plays the role of a hidden cost of otherwise inexpensive PV technology.

An established way of observing lateral nonuniformities in PV modules is infrared (IR) camera mapping: IR pictures of different parts of a module showing different temperatures; these nonuniformities depend on the device current and voltage. There exists a variety of complimentary mapping techniques, which also reveal lateral nonuniformities: optical beam induced current (OBIC), electron beam induced current (EBIC), surface photovoltage (SPV) mapping, photoluminescence (PL) mapping, and some others (reviews are given in Refs. \onlinecite{K1,K2}).

While the interpretation of some details of these maps requires further insights, the consensus in the PV community is that the observed lateral nonuniformities are related to material/structure imperfections or nonuniform light distribution. (As an example, we have polled a number of PV scientists on the nature of IR detected hot spots: all the responses attributed them to defects.) This understanding has matured to the level of common belief. Even the observed fact that the IR hot spots are typically close to the module bus bars was interpreted as an evidence of the bus bar application creating defects in the device. This understanding of hot spots has been documented in multiple publications, Refs. \onlinecite{kim2013,spanoche2013,qasem2013} presenting recent examples.

This paper will shift the defect paradigm towards spontaneous hot spot formation in a perfectly uniform system. Such hot spots appear as a result of runaway instability related to the diode-like current voltage characteristics of the device: current hogging by warmer regions makes them still warmer. \cite{karpov2013}

The phenomenon of current hogging belongs to a large class of runaway instabilities known in electrical engineering, chemistry (exothermic reactions), astrophysics (runaway nuclear fusion), where increase in temperature causes positive feedback. Its understanding is most advanced for thermal explosions. \cite{frank1969,zeldovich1985,kotoyori2005} In particular, it was realized that the instability starts with a hot spot \cite{merzhanov1971,thomas1973} resembling nucleation processes in phase transitions of the first kind; this analogy was explored in Ref. \onlinecite{subashiev1987} and then discussed for thin film structures. \cite{karpov2012}

Regardless of being spontaneous, the hot spot nonuniformity deteriorates device performances as different parts of the device operate under different temperatures and cannot be optimized simultaneously.

Furthermore, one can expect that hot spots will generate defects at exponentially higher rates compared to the surrounding cold regions. This will reverse the cause and effect relation between the defects and hot spots, the latter becoming cause rather than effect. In a long run, this scenario (mostly beyond the scope of this paper) will result in accumulation of permanent defects in certain local regions of PV modules and their corresponding degradation.

This paper is organized as follows. Section \ref{sec:exp} presents our experimental finding on hot spot formation and evolution. The algorithm and results of extensive numerical modeling of hot spots in thin film PV are described in Sec. \ref{sec:mod}. We then present a simplified analytical model of hot spot formation in Sec. \ref{sec:appr}. General discussion and conclusions are given in Sec. \ref{sec:concl}.

\section{Experimental data}\label{sec:exp}

\begin{figure}[htb]
\includegraphics[width=0.45\textwidth]{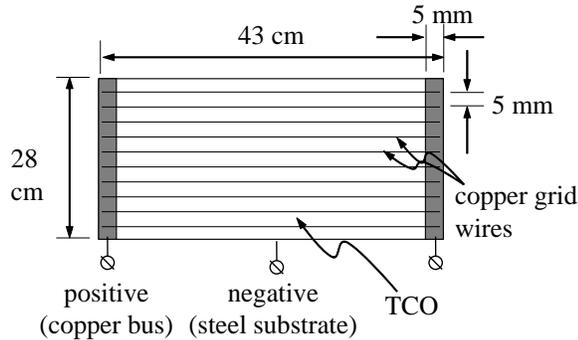}
\caption{Schematic of the cell used in our experiments; not to scale. Dark strips represent bus bars electrically disconnected from the transparent conductive oxide (TCO).\label{Fig:module}}
\end{figure}

\begin{figure}[htb]
\includegraphics[width=0.35\textwidth]{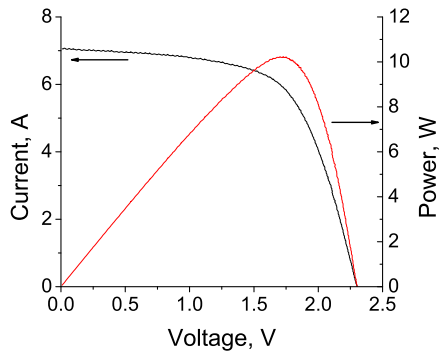}
\caption{The typical current voltage and power voltage characteristics of the studied cells.\label{Fig:IV}}
\end{figure}

We conducted electrical measurements and infrared temperature mapping on commercially available Xunlight Corporation manufactured photovoltaic cells. The cells are made of triple junction a-Si:H based structures deposited on 0.125 mm thick stainless steel, which acts both as mechanical support as well as the negative contact of the cell. The front of the cell is covered with a grid of copper wires bonded to the surface with a conductive adhesive. The diameter of the copper wires is 0.125 mm and the spacing between the wires is 5 mm. The cells are 43 cm x 28 cm in size.  A schematic of the cell and typical IV characteristics are shown in Figs. \ref{Fig:module} and \ref{Fig:IV}.

Unencapsulated cells were used for the purpose of the test, although the effects we describe may be observed with encapsulated cells as well. The cells were forward biased with a constant current power supply. A calibrated thermal imaging camera was used to map surface temperatures periodically.

During preliminary testing it was determined that the development of hot spots was sensitive to convective air currents, but the development of these currents was unpredictable and led to results that were not repeatable. Covering the cells to control convection was effective but interfered with the thermal imaging. Therefore we left the cells uncovered but directed a constant air stream from an electric fan on to the test area in order to provide a more consistent thermal environment during data collection.

\begin{figure}[h!]
\includegraphics[width=0.38\textwidth]{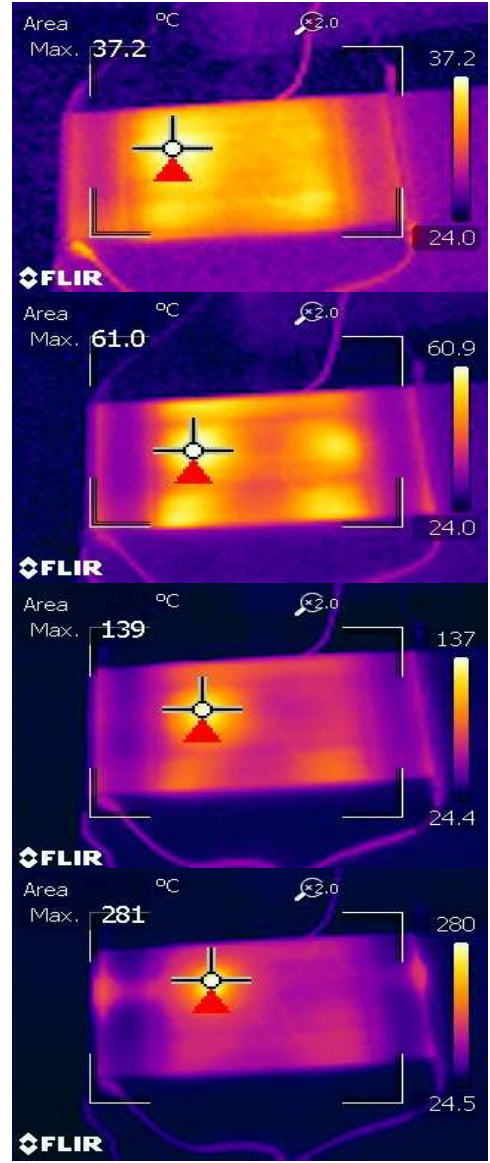}
\caption{IR mapping of a PV module showing the hot spot development and its temperature increase. The left side bar is under positive voltage and grids are parallel to the long dimension of the module. The current forced through the entire module was 14 A. \label{Fig:hotspot1}}
\end{figure}

Cells to be tested were placed manually in the test area and electrical connections were made. The cells were then left undisturbed for approximately 30 minutes to allow temperatures to equalize across the surface. The thermal camera was switched on and current was then forced in a forward direction through the cell. For all measured modules the initial uniform temperature
distribution became less uniform in the course of heating, and hot spots usually developed in the proximity of bus bars as illustrated in Fig. \ref{Fig:hotspot1}.

\begin{figure}[ht]
\includegraphics[width=0.32\textwidth]{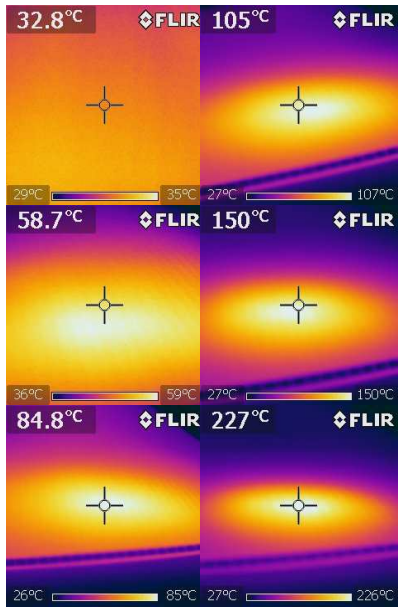}
\caption{The evolution of temperature distribution in a 10x10 cm$^2$ region close to the positive bus bar. \label{Fig:hotspot2}}
\end{figure}

\begin{figure}[ht]
\includegraphics[width=0.43\textwidth]{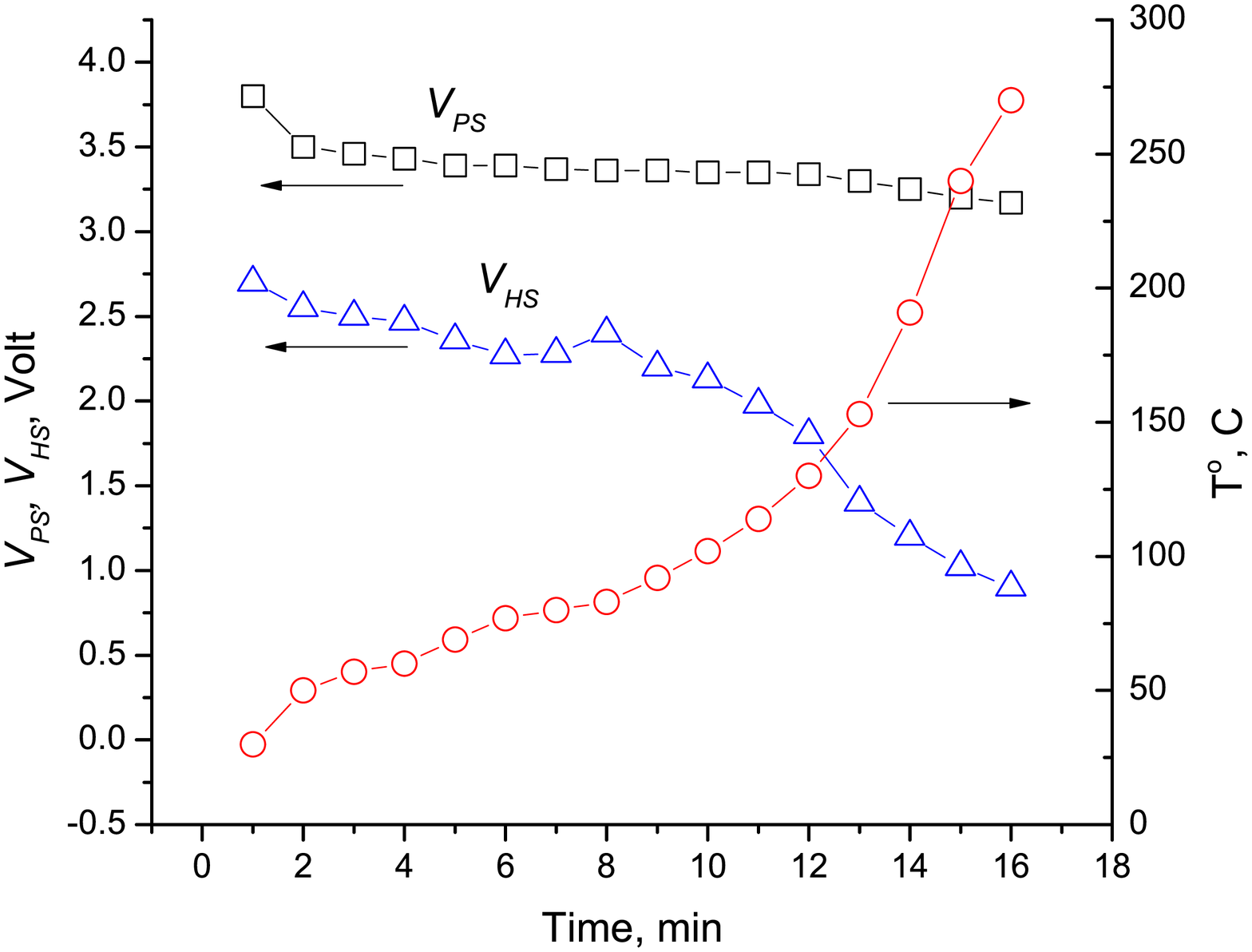}
\caption{Temporal variations of voltage on the power source ($V_{PS}$, voltage in the center of a hot spot ($V_{HS}$), and the hot spot maximum temperature. \label{Fig:VT}}
\end{figure}
The data in Fig. \ref{Fig:hotspot2} focuses on a single spot development in a relatively small localized region ignoring the rest of module area. It shows again a very considerable temperature increase and concomitant localization of its corresponding (hot spot) area.

We have observed that the hot spot formation takes place under forward bias conditions when the forward current exceeds certain critical value $I_c\approx 14-16$ A slightly varying between nominally identical devices.

Fig. \ref{Fig:VT} represents the typical data on the hot spot voltage and maximum temperature. We note that the voltage $V_{PS}$ on the power source slightly varies in the beginning of the process. This low temperature transient could be due to the initial uniform heating and some capacitive processes that are not related to the hot spot formation. $V_{PS}$ remains practically constant for the rest of the process where hot spots appear and evolve to their final shape. Both the voltage $V_{HS}$ and the hot spot maximum temperature are superlinear in time, although they vary smoothly and their rates of change are correlated.

An important feature beyond Fig. \ref{Fig:VT} is that the hot spot temperature and voltage show the trend of saturation after $\sim 20$ min; in particular, the temperature approaches $\approx 300$ $^o$C. Unfortunately, this saturation effect could not be studied more in detail, since maintaining such a high temperature for a considerable time resulted in irreversible changes, such as local discoloration and some others. These permanent device changes fall beyond the present scope. Here and in what follows we show only the data below the range of irreversible changes.

\section{Numerical modeling}\label{sec:mod}

\subsection{Electrical Model}\label{sec:emodel}

A numerical model of the device was developed to simulate its electrical and thermal properties.

The device to be modeled is a large-area (28 cm x 43 cm) a-Si based module.  The substrate and negative contact is made of a sheet of steel.  The steel components are responsible for the heat conduction within the plane of the device and for the bulk of the device's thermal mass; they are modeled with zero sheet resistance layers.

The positive contact of the device is a transparent conducting oxide (TCO) of relatively high sheet resistance ($\sim 150 \Omega /\square$ for the case under consideration), overlaid with regularly spaced copper grid lines ($\sim 10^{-4}$ $\Omega /\square$). The bus bars on the edges are parallel to the short dimension as illustrated in Fig. \ref{Fig:module}.
%Due to its size and complexity, some simplifications in modeling the grid are used in what follows.

\begin{figure}
\includegraphics[width=0.4\textwidth]{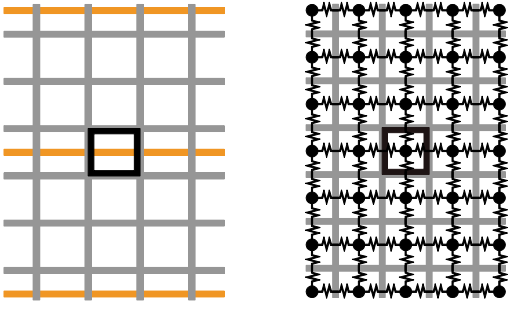}
\caption{Left, physical grid and TCO structure divided into nodes. Right, model of grid and TCO as nodes connected by discrete resistors \label{Fig:grids2}}
\end{figure}

In our modeling, the device is described through many points at the centers of their corresponding square nodes as shown in Fig.  \ref{Fig:grids2}. Each point is connected by a resistor to the edge of the node where it, in turn, is connected to a resistor of the adjacent node, as shown in Fig. \ref{Fig:grids2}, right.  For a node which only contains TCO and not a grid, all the associated resistors equal half the sheet resistance of the TCO.  Since two resistors are between each node, this choice implies that the total resistance between two points on the TCO equals the total TCO sheet resistance.

\begin{figure}
\includegraphics[width=0.38\textwidth]{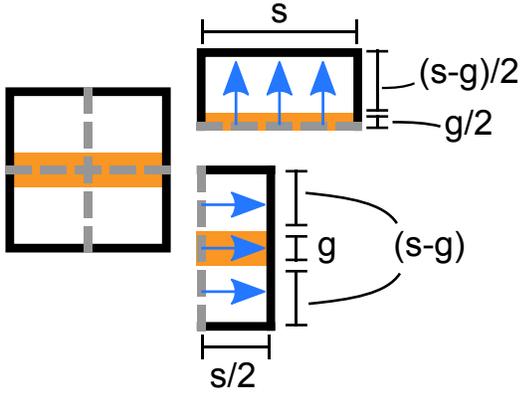}
\caption{Left, node containing grid and TCO. Upper right, geometry for determining resistors perpendicular to the grid.  Lower right, geometry for determining resistors parallel to grid. \label{Fig:rcalc}}
\end{figure}

A node containing a grid will also contain TCO, and the resistors in the direction parallel to the grid will differ from the resistors in a direction perpendicular to the grid. Fig. \ref{Fig:rcalc} defines the resistors in a node of length side $s$ containing a grid line of width $g$.  Resistors perpendicular to the grid line (in the vertical direction, as depicted) have value equal to the resistance across a rectangle, from the horizontal dividing line of the node to a horizontal edge of the node.  This is equivalent to resistors in series: the grid plus TCO material.  The resistance across a rectangle is the sheet resistance multiplied by the aspect ratio of the rectangle.  Consequently, if the grid material has sheet resistance $R_{grid}$, and the TCO has grid resistance $R_{TCO}$, then the equivalent resistance of a resistor perpendicular to the grid is
\begin{equation}
R_{\perp}=\frac{g}{2s}R_{grid}+\frac{s-g}{2s}R_{TCO}.\label{eq:Rper}\end{equation}
For resistors parallel to the grid, the resistance is across a rectangle, from the vertical dividing line of the node to a vertical edge of the node.  The resistances here are in parallel; hence, the total parallel resistance obeys the relation
\begin{equation}
\frac{1}{R_{\parallel}}=\frac{2g}{s}\frac{1}{R_{grid}}+\frac{2(s-g)}{s}\frac{1}{R_{TCO}}.\label{eq:Rpar}\end{equation}

\begin{figure}
\includegraphics[width=0.25\textwidth]{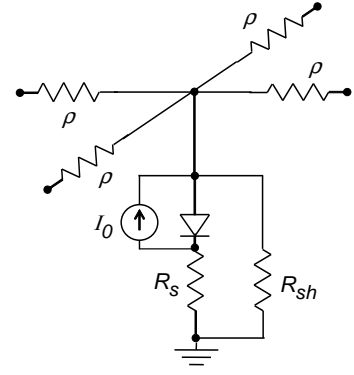}
\caption{The equivalent circuit of a node showing TCO/grid structure on top, and local active photovoltaic structure. \label{Fig:activel}}
\end{figure}

The preceding describes the numerical model of the negative contact and grid structure.  The active solar cell material between the TCO and steel substrate is modeled as shown in Figure \ref{Fig:activel}.

Although the device is in fact a triple junction amorphous silicon cell, for simplicity the active layer is modeled as a single diode. The diode is modeled with the standard diode current voltage (IV) characteristic,
\begin{equation}
I=I_{0} \left[\exp\left( \frac{qV}{nk_{B}T} \right) -1 \right] \quad I_{0}=I_{00} \exp\left( -\frac{E}{k_{B}T} \right)\label{eq:Shockley}\end{equation}

Here, $q$ is the electron charge, $n$ is the diode ideality factor, $k$ is Boltzmann's constant, $T$ is the local temperature.  $I_{0}$, the reverse saturation current, is found through the cell known open circuit voltage $V_{OC}$ and short circuit current density, $j_{L}$.  If $s^2$ is the node area, then the equation
\begin{equation}
j_{L}=I_{0}\left[\exp\left( \frac{qV_{OC}}{nkT}\right) -1  \right]/s^{2}   \label{eq:Iodef}\end{equation}
enables one to relate the above parameters to the experimental data. We used $j_{L} = 5$ mA/cm$^2$ and $V_{OC} = 2.2V$. We then chose $n = 6$ (as though each junction in the triple stack had $n = 2$) to provide satisfactory fits of the measured IV curves. The activation energy $E$ was determined by the experimentally known variation in open circuit voltage with temperature $T$ that is $0.38$ $\%/$K near room temperature yielding $En=4.68$ eV.

While the IV fitting becomes possible with Eq. (\ref{eq:Iodef}), the latter equation still represents a significant simplification neglecting possible contributions due to the series and shunt resistances of a {\it non-ideal} diode (see e. g. Ref. \onlinecite{sze}). These effects may become visible especially under the conditions when the latter resistances are temperature dependent, which certainly takes place with semiconductor devices. We will see in Sec. \ref{sec:modelres} that these effects do create a discrepancy between the modeling and experimental data. We have verified that including these effects in our modeling improves the agreement and can be easily done when needed. However, it also brings in several new adjustable parameters and makes the model over-flexible. Therefore, aiming at understanding the basic physics of the hot spot phenomena we have limited this study to a simple IV characteristics in Eq. (\ref{eq:Iodef}).

The active layer was modeled as including both ideal diodes (obeying the Shockley equation) in connection with parasitic series and shunt resistances, $R_{s}$ and $R_{sh}$, respectively.  When the device is laterally uniform, the products of series or shunt resistance times device area, which we call the specific series and shunt resistances, $R_{s}^{\prime}$ and $R_{sh}^{\prime}$, are constant.  Therefore,
\begin{equation}
R_{s}= \frac{R_{s}^{\prime}}{ s^2},\quad R_{sh}= \frac{R_{sh}^{\prime}}{ s^2}  \label{eq:rsdef}\end{equation}

Having defined the model elements, the voltage distribution over the device (voltages at each node) is found.  This is done through Kirchhoff's law that the sum of currents entering a node must equal zero,
\begin{eqnarray}\label{eq:kirch1}
&&\frac{V(x+s,y)-V(x,y)}{\rho (x+s,y) + \rho (x,y)}+\frac{V(x- s,y)-V(x,y)}{ \rho (x- s,y) + \rho (x,y)}+
  \nonumber \\
&&\frac{V(x,y+s)-V(x,y)}{\rho (x,y+s) + \rho (x,y)} +
\frac{V(x,y-s)-V(x,y)}{\rho (x,y-s) + \rho (x,y)} =\nonumber \\
&& \frac{V(x,y)}{R_{sh}^{\prime}}s^2+ I_{D}(x,y)
\end{eqnarray}
Here $I_{D}$ is the current through the diode and $\rho$ is the generalized sheet resistance that can represent $R_{perp}$ and $R_{parallel}$ from Eqs. (\ref{eq:Rper}) and (\ref{eq:Rpar}) respectively.

For nodes on the edge of the device, for which nodes at $x \pm s$, $y\pm s$ do not exist, the corresponding terms involving those coordinates are not included.  This satisfies the boundary condition on the edges (no current flows into or out of those edges).  One other boundary condition imposed by the experimental setup is that $I_{bb}=16$ A of current were supplied to a bus bar.  As the latter has very low resistance, we have assumed the injection of this current distributed evenly throughout the busbar.
This condition (which we define to apply to the x=0 boundary) adds the following equations for all $y$s
\begin{equation}
\frac{L}{s}\frac{V(0,y)-V(s,y)}{R_{bb}} = I_{bb}
\label{eq:busbar}
\end{equation}
where $R_{bb}$ is the bus bar sheet resistance.

The above equations, over all $x$ and $y$, define $V(x,y)$.  For the system of spatially close diodes they can be linearized,
\begin{equation}
I_{D}(x,y) = I^{0}_{D} + \frac{\partial I_{D}}{\partial V}(V(x,y)-V^{0}(x,y))
\label{eq:lineari}
\end{equation}
where $V^{0}$ is the trial voltage at the node (its value may come from a guess or a previous iteration). Likewise, $I^{0}_{D}$ is the constant current through the diode when the node voltage is $V^{0}$.  We have used iterative techniques to solve the above linear system.

More specifically, the linearization was applied as follows.  First, as the diode and resistor $R_S$ are in series, the same current flows through both.  If $V_n$ is the voltage at the node between the diode and the resistor, then this current obeys the equation
\begin{equation}
I_{D}=I_{0} \left[\exp\left(\frac{q(V(x,y)-V_{n})}{nkT}\right) -1\right]=\frac{V_n}{R_S}.
\label{eq:lineari2}
\end{equation}
Its solution can be expressed through the Lambert $W$ function defined as $W(x)\exp[W(x)]=x$. The result takes the form
\begin{equation}
I_D = -I_0 + \frac{nkT}{qR_S}W( \theta),
\end{equation}
\begin{equation}
\frac{\partial I_D}{\partial V}=\frac{qI_0}{nkT}\frac{W( \theta)}{1+W( \theta)}\exp\left[\frac{q(V(x,y)+I_0 R_S}{nkT}\right]
\end{equation}
where we have introduced the new variable,
\begin{equation}
\theta = \frac{qI_0 R_S}{nkT} \exp\left[ \frac{(V(x,y)+I_0 R_S}{nkT/q}\right].
\end{equation}

The node size is determined by a compromise between computational time (using fewer, larger nodes) and accuracy (using more, smaller nodes).  For the voltage distribution to be qualitatively accurate, it was necessary for there to be at least two nodes between adjacent grid lines. The simulation accuracy was checked by using higher resolution and verifying that the results had converged.

\subsection{Thermal Model}\label{sec:tmodel}

The time evolution of temperature in the device is governed by the heat equation
\begin{equation}\label{eq:thermal}C \partial T/\partial t = Q -\alpha (T-T_0)+\nabla (\kappa \nabla T)-\sigma (T^4-T_0^4).\end{equation}
Here $\alpha$ is the thermal exchange coefficient in the Newton law of cooling and $T_0$ is the ambient temperature assumed constant along the module surface, and $\sigma\approx 5.67\times 10^{-8}$ Wm$^{-2}$K$^{-1}$ is the Stefan-Boltzmann coefficient in the law of radiation cooling; all other variables have their standard meaning. For a thin sheet of thickness $d$, it is more convenient to use two dimensional description assuming $T$ constant along the transversal direction. The radiation cooling may be significant at high enough spot saturation temperatures approaching $T_{sat}\approx 600$ K in our observations.

The discrete version of the heat transfer equation takes the form
\begin{eqnarray}\label{eq:heatt}
&C\partial T/\partial t = Q(x,y) - \alpha (T-T_{0})-\sigma (T^4-T_0^4)\nonumber \\&-\chi [4T-T(x+s,y)-
T(x-s,y)\nonumber \\ &-T(x,y+s)-T(x,y-s)]
\end{eqnarray}
where $T\equiv T(x,y)$, $C$ and $Q$ are the heat capacity and heat generated within one node, and $\chi = \kappa d$ is a thermal sheet conductivity independent of node size. For the simulations presented, we have used $\kappa = 16$ Wm$^{-1}$K$^{-1}$,  $C=0.466$ Jg$^{-1}$K$^{-1}$ for the stainless still of mass density 7.9 g/cm$^3$ and $d = 125$ $\mu$m. $Q$ is the sum of Joule heating terms IV for all resistive elements and the diode within the node. $\alpha$ the heat transfer coefficient (which we took to be that of air against mild steel, $\alpha =8$ Wm$^{-2}$K$^{-1}$) multiplied by the node area, $s^2$.

When k is not constant with position, the above $\chi$ should be replaced with the effective value
\begin{equation}
\chi _{eff} = 2\left[ \frac{1}{ \chi (x,y)}+\frac{1}{ \chi (x \pm s, y \pm s)}\right]^{-1}
\end{equation}
where the argument of the second $\chi$ term must be the same as the corresponding temperature term $T(x \pm s, y \pm s)$ it multiplies.  This can be seen  by considering $1 / \chi _{eff}$ to be a sum of the two thermal resistances, $1 / \chi(x,y)/2$ and $1 / \chi(x \pm s, y \pm s)/2$, added in series.  As with the voltage model, if nodes at $x \pm s$, $y\pm s$ do not exist within the device, the corresponding differences [$T-T(x \pm s, y \pm s)$] involving them must be neglected.

The algorithm of thermal modeling includes the following three steps.  (i) Determine the voltage distribution over the device. (ii) Find the local heat generated within each node from the voltage distribution. (iii) Integrate in time with Eq. (\ref{eq:heatt}). The algorithm can be optimized by addressing the following issues: That step (i) is relatively computationally expensive, and additionally, step (iii) has possibilities of stiffness.

We mitigate the stiffness by noting that (the continuous counterpart of) Eq. (\ref{eq:heatt}) can be put in the form
\begin{equation}
\frac{\partial T}{\partial t}(x,y,t)=A - B \cdot T(x,y,t)
\end{equation}
Assuming $A$ and $B$ are constants, this has solution
\begin{equation}
T(x,y,t_0+\delta)=\frac{A}{B}+\left ( T(x,y,t_0)-\frac{A}{B} \right )\exp (-B \delta ).
\label{eq:tstep}
\end{equation}
Taking time steps of this form, as opposed to that by Euler or Runge-Kutta recipes, prevents divergence and spurious oscillation regardless of step size, and also has the physical property that no spot on the device can ever have temperature below $T_{0}$.  Assumed in deriving this is that $A$ is a constant.  In reality, $A$ is made of an internal heat term as well as terms depending on adjacent temperatures.  While we may assume the heat does not change greatly as a function of time, the temperature of adjacent nodes in the device will change approximately as fast as in the node under consideration.  This can be dealt with by using small step sizes, which is not a limitation as this integration step is not slow.

As computing the voltage distribution is time consuming, we wish to assume it is held constant for larger periods of time $\Delta >> \delta$.  To check the validity of this assumption and to improve accuracy, from time $t_0$ we make steps of the form in Eq. (\ref{eq:tstep}) to time $t_{o}+\Delta /2$.  At this time, the voltage distribution is found again, as well as the local heat generation.  We then return to time $t_0$ and, using the heat generation found at time $t_{0}+\Delta /2$, step to time $t_{0}+\Delta$.  As this passes $t_{0}+\Delta /2$ a second time, it is possible to compare the temperatures obtained with the two constant heats at these times.  If the difference becomes too large (e. g., 0.5K), it is possible to adaptively decrease the step size.

\subsection{Modeling Results}\label{sec:modelres}

\begin{figure}[t!]
\includegraphics[width=0.50\textwidth]{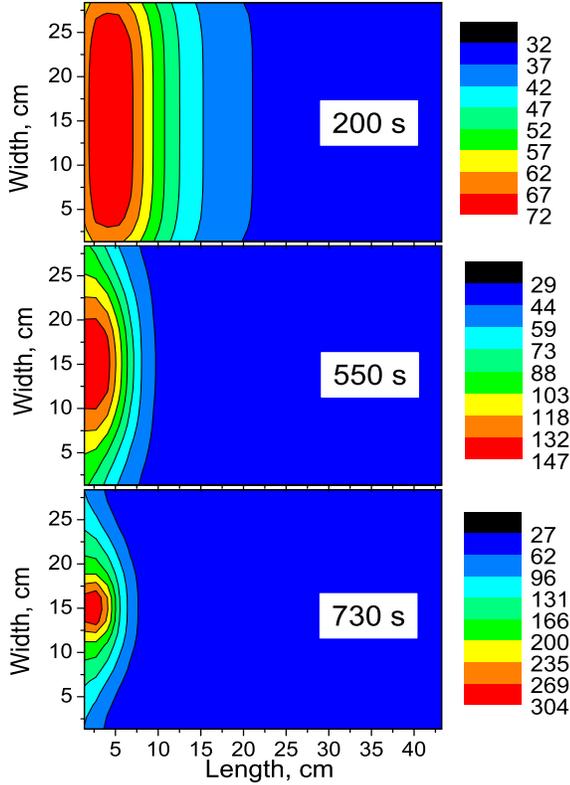}
\caption{Simulated temperature distributions at different time instances. Note the differences in temperature scales shown to the right.  \label{Fig:tmap}}
\end{figure}

\begin{figure}[t!]
\includegraphics[width=0.40\textwidth]{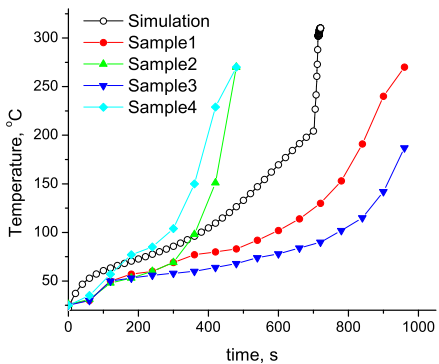}
\caption{Simulated and experimental temporal dependencies of temperature at the hottest spot in a cell.  \label{Fig:tspot}}
\end{figure}

\begin{figure}[t!]
\includegraphics[width=0.40\textwidth]{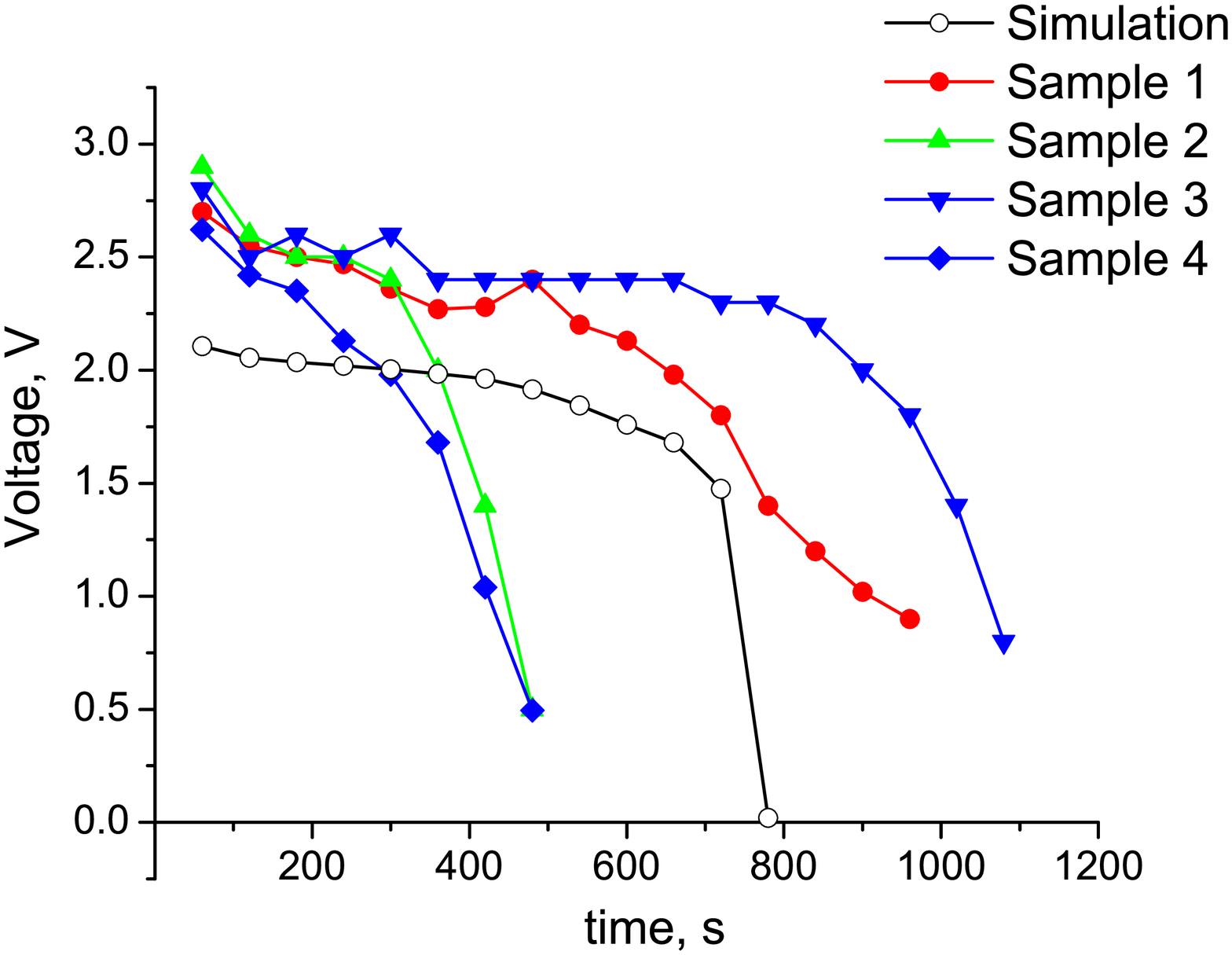}
\caption{Simulated and experimental temporal dependencies of voltages at the hottest spot in a cell.  \label{Fig:vspot}}
\end{figure}

The above modeling has allowed us to simulate the temperature distribution over the sample (Fig. \ref{Fig:tmap}), as well as the temporal evolution of hot spot temperature (Fig. \ref{Fig:tspot}) and voltage (Fig. \ref{Fig:vspot}). These and subsequent results are limited to low enough temperatures below $600$ K where experimentally the system does not show significant irreversible changes. We note that the experimentally observed saturation temperature close to 600 K is reproduced by our modeling when the radiation cooling is taken into account, while neglecting that mechanism results in simulated saturation temperatures of $\gtrsim 1000$ K inconsistent with the data. We note that, from the modeling results, the temperature rolling over to saturation region is relatively short, $\sim 10$ s, after which the temporal dependence becomes almost flat; the beginning of that region is shown in Fig. \ref{Fig:tspot}.

The temperature maps in Fig. \ref{Fig:tmap} clearly demonstrate the phenomenon of runaway instability where the spot gets hotter and simultaneously shrinks in its linear dimensions; the temperature of far away regions simultaneously decreases as they dissipate smaller currents (the total current remains fixed). The trends in the simulated data mirror those in the experimental data, and the spatial scale of the temperature nonuniformity is also similar to that observed in experiment, for comparable sample temperatures.

Figure \ref{Fig:tspot} shows the hottest temperature over the sample versus time after power is applied, comparing simulated data to data obtained from different experimental samples.  Generally, the simulated curve has behavior similar to that of the data, indicating that the simulation captures most of the important details.  We note that even nominally identical real cells can enter thermal runaway at different times. In simulations, we have found that the series resistance parameter can control the time to runaway, with variations on the order of $50\%$ being sufficient to cover experimentally observed times to runaway.  Although the simulated data presented here is for a spatially uniform device, simulated devices with existing spots of locally low series or shunt resistance (including spots of low shunt resistance that nevertheless have negligible effect on device current-voltage and performance properties) can influence time-to-runaway and hot spot location.  Another possible effect which could be included in future simulations, though neglected for the presented data, could be a more detailed model of the temperature dependence of the series resistance and active layers, as discussed in the above, after Eq. (\ref{eq:Iodef}). This is also seen from Fig. \ref{Fig:vspot} where the neglected effect of temperature dependent series resistance is responsible for the voltage difference between the simulation and the experiment.

\begin{figure}
\includegraphics[width=0.38\textwidth]{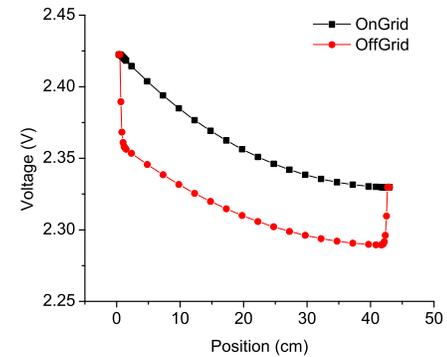}
\caption{Initial voltage distribution along the length of the cell, on and off the grid. \label{Fig:voltini}}
\end{figure}

\begin{figure}
\includegraphics[width=0.38\textwidth]{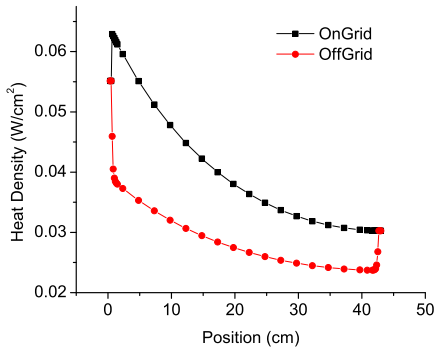}
\caption{Initial heat generation within the cell. \label{Fig:heatini}}
\end{figure}

More in detail, the development of the hot spot can be understood from our modeling results as follows.  Fig. \ref{Fig:voltini} shows the initial voltage on the middle of the sample, parallel to the grid lines, on and off a grid line.  The corresponding heat generation density is presented in Fig.  \ref{Fig:heatini}. The generated heat is initially greatest at the grid lines, nearest the busbars.  This is due to the active layer; the contribution due to resistive losses in the bus bar and grids is relatively low. The generated heat is greatest there for two reasons: The current density is highest there, due to it being more concentrated than in the bus bar, and the branching current through the active layer is greater on the grid than the bus bar due to the sheet resistance of the former being greater. Although there is yet more branching current through the area of the cell covered by TCO only, the corresponding voltage there is also lower, and so is the generated heat. Consequently, the region that initially heats fastest is close to but off of the busbar. An additional feature that moves the initial hot spot away from the bus bar is that its material adds to the local heat capacity and thermal conductivity.

\begin{figure}
\includegraphics[width=0.4\textwidth]{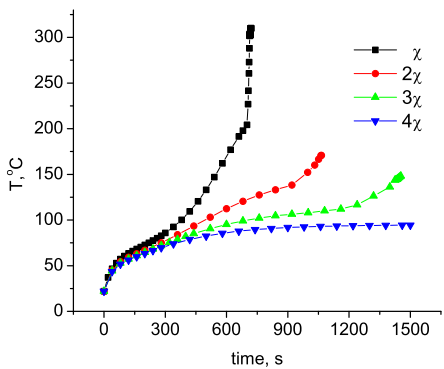}
\caption{Simulated temporal temperature dependencies as a function of substrate thermal conductivity. \label{Fig:chitemp}}
\end{figure}

It follows from the above that, as a consequence of the voltage gradients and nonuniform heat generation, temperature nonuniformity over the sample is inevitable.  It is practically important however that runaway is not inevitable.   Fig. \ref{Fig:chitemp} shows simulated temperature vs. time curves for cells with identical properties as the simulated cell referenced in Fig. \ref{Fig:tspot}, except that the substrate thermal conductivity ($\chi$) has been increased by a factor.  With $\chi\rightarrow 4\chi$ (achievable in practice), the cell never enters runaway, and reaches a maximum temperature of $90^{\circ}$C.

Finally, we note that the threshold nature of runaway instability was reproduced by the above modeling as well. The chosen set of parameters led to the prediction of runaway instability starting above the total current of approximately 14 A, close to the experimentally observed values.

\section{Approximate analytical model}\label{sec:appr}
\begin{figure}[t!]
\includegraphics[width=0.40\textwidth]{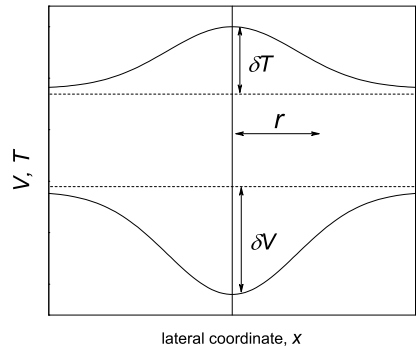}
\caption{Sketches of temperature and electric potential distributions across the hot spot.  \label{Fig:sketch}}
\end{figure}

As before, a hot spot is characterized by its related temperature and electric potential distributions illustrated in Fig. \ref{Fig:sketch} where $\delta T\equiv T-T_0$ and  $\delta V\equiv V_0-V$ where $T_0$ and $V_0$ are the temperature and potential far from the spot. For simplicity, we assume that $T_0$ coincides with the ambient temperature.

The actual coordinate dependencies of $\delta T$ and $\delta V$ are described by a system of coupled nonlinear differential equations
\begin{eqnarray}\label{eq:distr}
\nabla ^2V&=& \rho j(V,T), \\
c\frac{\partial T}{\partial t} &=& H+jV+\chi \nabla ^2T-\alpha (T-T_0)-\sigma (T^4-T_0^4)\nonumber.
\end{eqnarray}
The former describes current branching through the diode elements between the resistive and ground (ideal) conductor; \cite{Luk,Kar2} the latter describes the heat transfer. Here $\nabla ^2$ stands for the two-dimensional Laplacian in the lateral directions, $\rho$ is the sheet resistance,  $H$ is the uniform heat per area, $c$ is the specific heat per area, and $\chi \sim h\chi _0$ is the "sheet" thermal conduction the dimensionality of W/K (cf. Sec. \ref{sec:tmodel}). The diode-type current voltage characteristic expresses current density (A/m$^2$), e. g.
\begin{equation}\label{eq:jV}
j=j_{00}\exp(-E/kT)[\exp(qV/kT)-1]-j_L,
\end{equation}
or similar including shunt and series resistances.

The linear analysis of Eqs. (\ref{eq:distr}) does predict the threshold nature of runaway instability. \cite{karpov2013} However, solving these equations beyond the linear approximation is a challenging mathematical problem that we will not pursue in this work.

Here, we limit our selves to a simplified approach using the following rough approximations.\\
1) The one-parameter scaling,
\begin{equation}\label{eq:onepar}
T(x)=T_0+\delta Tf_T(x/r),\quad  V(x)=V_0+\delta Vf_V(x/r)
\end{equation}
where $r$ is the characteristic radius of the spot, and undefined functions  $f_T(x/r)$ and $f_V(x/r)$ both equal unity at $x=0$ and zero at $x\rightarrow \infty$. In this approximation, $\nabla ^2 T\sim \delta T/r^2$, $\int (V_0-V) xdx\sim r^2\delta V $, etc.\\
2) For the actual values of parameters, the left hand side $C\partial T/\partial t$ in Eq. (\ref{eq:distr}), is relatively small; the terms in the equation right hand side almost cancel each other. That is the quasistatic approximation with $C\partial T/\partial t=0$.\\
3) For the saturated temperature, the thermal conduction term $\chi \nabla ^2T\sim \delta T/r^2$ must be comparable to the cooing terms $\alpha \delta T+\sigma (T^4-T_0^4)$; otherwise the hot spot could collapse to zero radius.\\
4) We approximate $T^4-T_0^4=4T_0^3\delta T$.

Under forward bias conditions,  Eqs. (\ref{eq:distr}) can be then approximately replaced with more intuitive equations,
\begin{eqnarray}\label{eq:Icur}
&(\alpha _{eff} +\chi /r^2)\delta T=Vj,\quad \alpha _{eff}=\alpha +4\sigma T_0^3, \nonumber \\& I=\delta V/\rho=\pi r^2j,\quad j=j_{00}\exp[-(E-V)/kT].
\end{eqnarray}

Eqs. (\ref{eq:Icur}) lead to the following results where the current $I$ must be considered as fixed by the external power source. Far from saturation, the hot spot radius is given by
\begin{equation}\label{eq:rdT}
r=\sqrt{\frac{IV}{\pi\alpha \delta T}} \quad (\gg r_{sat}).
\end{equation}
The ultimately small saturated radius is
\begin{equation}\label{eq:rsat}
r_{sat}=\sqrt{\chi/\alpha_{eff}} \sim 1\quad {\rm cm}.\end{equation}
The saturated temperature increase becomes
\begin{equation}\label{eq:Tsat}
\delta T_{sat}=VI/(2\pi \chi)\sim 1000\quad {\rm K}.\end{equation}
Finally,
\begin{equation}\label{eq:VT}
V=\frac{E}{q}-\frac{kT}{q}{\cal L}, \quad {\cal L}\equiv\ln\left[\frac{\pi\alpha _{eff}(T-T_0)}{j_{00}}\right].\end{equation}
All these results are qualitatively consistent with the experimental data and with the results of modeling above. Given the very rough nature of the underlying approximation, the agreement appears quite satisfactory; cf. the observed $r_{sat}\sim 2-3$ cm, and $T_{sat}\approx 600$ K.

Eq. (\ref{eq:rdT}) describes how the spot radius decreases with the temperature. This prediction is consistent with the modeling results in Fig. \ref{Fig:tmap} and experimental results in Figs. \ref{Fig:hotspot1} and \ref{Fig:hotspot2}. Furthermore, the data in Fig. \ref{Fig:VT} enables one to estimate $dV/dT\sim 10^{-2}$ V/K; hence, ${\cal L}\sim 30$. Because the above introduced ${\cal L}$ is relatively close to the logarithmic parameter ${\cal L}\sim 50$ from Ref. \onlinecite{karpov2013} (they differ by $\ln [\pi\alpha _{eff} (T-T_0)/j_L]\sim 10$), we conclude that Eq. (\ref{eq:VT}) is approximately correct even quantitatively.

As a result, simple intuitive equations (\ref{eq:Icur}) can be considered adequate qualitative description of the hot spots in thin film devices under forward bias. In addition to the above, they explain, through the coefficient $\alpha _{eff}$, how changing the ambient air flow affects the hot spot formation, again in agreement with the experimental observations (see Sec. \ref{sec:exp} above).

\section{conclusions}\label{sec:concl}
We have shown experimentally and by numerical modeling that hot spots with significant temperature increase ($\sim 300$ K) can spontaneously emerge in laterally uniform thin film photovoltaics. This is probably the most important conclusion of our work.

Since hot spots often observed in PV are not necessarily related to defects or other imperfections making devices laterally nonuniform, care should be taken to optimize the device design in a way allowing to avoid the runaway instability underlying the hot spot formation. We have shown that simple steps (such as changing to a more thermally conducting substrate; see Fig. \ref{Fig:chitemp}) can suppress the hot spot formation.

As already mentioned in Sec. \ref{sec:intro} above, nonuniform material degradation accelerates at hot spots, i. e. an initial hot spot may then degrade in a runaway mode under more and more stress as it becomes progressively more shunting. The final result of such degradation will be roughly one shunt per the area of the characteristic linear dimension of lateral nonuniformity \cite{K1,K2} $L\sim 0.1-1$ cm. It is remarkable that problems with performance and reliability related to hot spot instability can be fixed by properly scaling the device thickness, substrate material, and thermal insulation.

Summarizing more specific conclusions of this work, the following can be stated.\\
1) Hot spots appear close to the device electrodes (bus bars) under significant enough forward current.\\
2) Hot spots evolve in such a way that their temperature increases while the electric potential and spot radius decrease.\\
3) The thermal properties (specific heat, thermal conductivity, and Newton's cooling coefficient) are important parameters determining hot spot development.\\
4) The ambient temperature and thermal conduction are important as well: ambient cooling promotes thermal runaway. \\
5) Given the device structure, thermal runaway and its related hot spots can be numerically modeled thus allowing hot spot free device engineering. Numerical algorithm developed in this work or other algorithms \cite{lanz2013} can be used provided that they do not impose the often assumed restriction of device uniformity in the course of modeling.\\
6) Radiative cooling can be an important factor limiting the hot spot size and other parameters.\\
7) A semi-quantitative understanding of well developed hot spots can be achieved based on a system of intuitive simple equations.\\

One important aspect of hot spot formation left behind the present scope is a possible role of lateral nonuniformities always present in real devices. Based on the above developed understanding, one can assume that such nonuniformities can trigger the hot spot formation and their location; yet the final parameters of these spots will be determined by the average device parameters (except maybe some cases of extremely nonuniform structures). This conclusion follows from the above established fact that even in laterally uniform devices local spot heating is severe and stable enough to sustain the addition of faulty local elements.

\acknowledgements
This work was performed under the auspice of the NSF award No.  1066749.
Discussions with A. V. Subashiev and D. Shvydka, and experimental arrangements by P. Hildebrandt are greatly appreciated.

\end{document}